\documentclass[preprint,5p]{elsarticle}
\usepackage{graphicx}
\usepackage{dcolumn}
\usepackage{bm}  
\usepackage{amssymb}  
\usepackage{hyperref}

\hypersetup{colorlinks=true, urlcolor=blue, citecolor=blue}
\usepackage[displaymath]{lineno}
\usepackage{natbib}
\biboptions{sort&compress}

\title{\vspace{-15mm}\fontsize{18pt}{10pt}\selectfont\textbf{Rate capability 
and magnetic field tolerance measurements of fast timing microchannel plate 
photodetectors}}

\author[anl]{Junqi~Xie\corref{cor}}
\author[anl]{Mohammad~Hattawy}
\author[bnl]{Mickey~Chiu}
\author[anl]{Kawtar~Hafidi}
\author[anl]{Edward~May}
\author[anl]{Jose~Repond}
\author[anl]{Robert~Wagner}
\author[anl]{Lei~Xia}

\address[anl]{Argonne National Laboratory, Argonne IL 60439, USA}
\address[bnl]{Brookhaven National Laboratory, 2 Center St., Upton, NY, 11973, USA}
\cortext[cor]{Corresponding author: jxie@anl.gov (J. Xie)}

\date{\today}

\begin{document}

\begin{abstract}
Microchannel plate photodetectors provide both picosecond time resolution and 
sub-millimeter position resolution, making them attractive photosensors for 
particle identification detectors of a future U.S. Electron Ion Collider. We 
have tested the rate capability and magnetic field tolerance of 6$\times$6 
cm$^{2}$ microchannel plate photodetectors fabricated at Argonne National 
Laboratory. The microchannel plate photodetector is designed as a low-cost 
all-glass vacuum package with a chevron pair stack of next-generation 
microchannel plates functionalized by atomic layer deposition. The rate 
capability test was performed using Fermilab's 120 GeV primary proton beam, and 
the magnetic field tolerance test was performed using a solenoid magnetic with 
tunable magnetic field strength up to 4 Tesla. The measured gain of the 
microchannel plate photodetector is stable up to 75 kHz/cm$^{2}$, and varies 
depending on the applied magnetic field strength and the rotation angle 
relative to the magnetic field direction.
\end{abstract}

\maketitle


\section{Introduction} \label{sec:level1}
The world's first Electron Ion Collider (EIC) \cite{EIC} has been recommended
in the 2015 Long Range Plan for Nuclear Science as the highest priority for a 
new facility construction in the U.S following the completion of the Facility 
for Rare Isotope Beams (FRIB) \cite{LRP}. This unique facility will explore 
several features of the strong force and QCD by mapping the parton content, 
quarks and gluons, of nucleons and nuclei up to uranium through processes such 
as semi-inclusive deep inelastic scattering and exclusive deeply virtual 
Compton scattering.  

Several detector concepts have been proposed and designed for the EIC at 
Argonne National Laboratory (ANL), Brookhaven National Laboratory (BNL), and 
Thomas Jefferson National Accelerator Facility (JLab). For all these EIC 
detector designs, excellent particle identification (PID)\\ ($e$/$\pi$/$K$/$p$) 
over a wide range of momentum is essential for the proposed measurements.  
Time-of-flight systems and imaging \u Cerenkov detectors, such as Ring Imaging 
\u Cerenkov (RICH) \cite{RICH,RICH2} and Detection of Internally Reflected \u 
Cerenkov light (DIRC) \cite{DRIC} detectors, are proposed. Both of these 
detector classes are calling for large area, low cost photon sensors with high 
spatial resolution, high rate capability, radiation tolerance, magnetic field 
tolerance, and picosecond timing resolution. 

Microchannel plate (MCP) photodetectors are compact photon sensors, usually 
with an internal chevron pair stack of MCPs. This  geometry provides both high 
spatial and temporal resolution in a vacuum package. The Large Area Picosecond 
Photodetector (LAPPD$^{TM}$) is the world's largest MCP based photodetector 
with an active area of 20$\times$20~cm$^2$ \cite{LAPPD}. It is designed as a 
modular all-glass detector package with the next-generation MCPs produced by 
applying resistive and emissive coatings to borosilicate glass capillary array 
(GCA) substrates through atomic layer deposition (ALD). The all-glass design 
and low-cost next-generation MCPs provide great advantages in reducing the 
LAPPD$^{TM}$ product cost per area compared to other currently available MCP 
based photodetectors. As a collaborating group of the LAPPD project 
\cite{LAPPD2}, we have built an MCP photodetector fabrication system 
\cite{LAPPD-ANL} at ANL to fabricate 6$\times$6~cm$^2$ MCP photodetectors with 
LAPPD design. The ANL's MCP photodetector fabrication system also serves as an 
R{\&}D platform for LAPPD package design validation and optimization.  Several 
6$\times$6~cm$^2$ MCP photodetectors with standard LAPPD design were 
successfully fabricated at ANL and tested \cite{ANL-MCPs,Wang-MCPs,Wang-MCPs2}, 
exhibiting a high gain over 10$^7$, an overall time resolution of 35 ps, and a 
position resolution better than 1~mm. The excellent performance of these 
6$\times$6~cm$^2$ MCP photodetectors shows that the low-cost LAPPD$^{TM}$ 
detector is a promising candidate photosensors for the EIC PID detector 
systems. Performance tests of the MCP photodetectors in high rate, high 
radiation damage, and high magnetic field environments are required to further 
validate the application of LAPPD$^{TM}$ detectors as the EIC PID photosensors. 

In this paper, we describe the current design of 6$\times$6~cm$^2$ MCP 
photodetectors fabricated at ANL, report recent results on their performance tests in a high 
rate environment and a high magnetic field environment. The direction of 
further optimization of the LAPPD$^{TM}$ design for an EIC PID application is 
also addressed at the end of this paper.

\begin{figure*}[tbp]
\centering \includegraphics[scale=0.3]{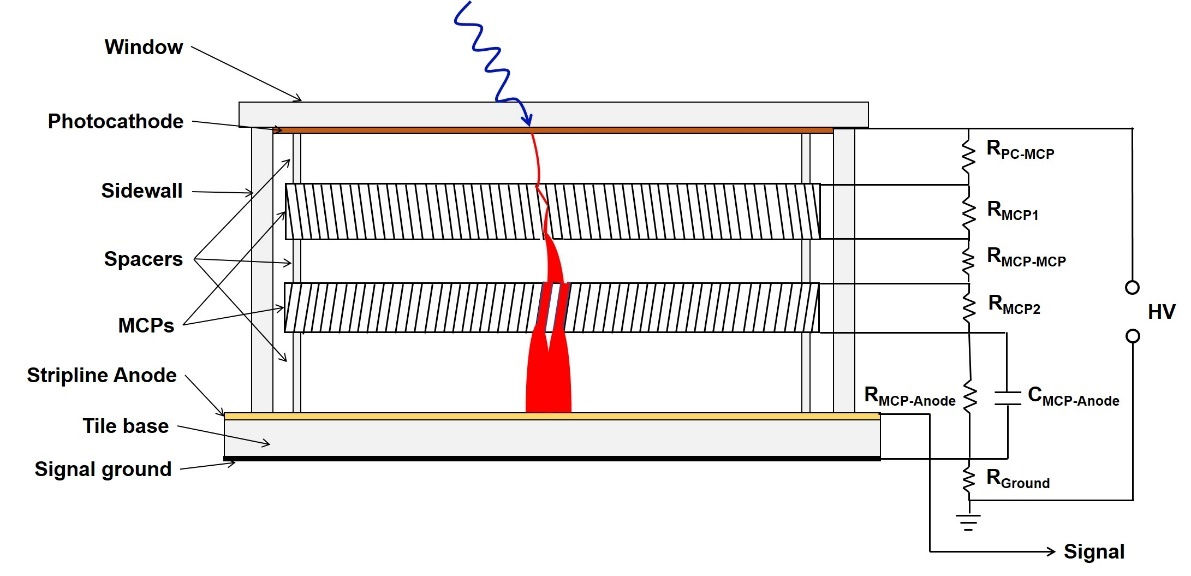}
\caption{Schematic of MCP photodetector assembly (not to scale) and the 
   electrical circuit diagram. External connections to the top and bottom 
surfaces of the two MCPs are through ultra-thin metal shims (not shown) to 
special extra striplines on the tile base. The circuit diagram shows 
connections through side wall in a simplified format.} \label{fig:design}
\end{figure*}

\section{Design of the MCP photodetector assembly} \label{sec_design}
The current ANL design of the 6$\times$6~cm$^2$ MCP photodetector is developed 
from the original LAPPD internal resistor chain design \cite{Wang-MCPs2}, 
similar to the current commercial standard design of LAPPD$^{TM}$ detectors 
\cite{Craven-MCPs}. Figure \ref{fig:design} shows a schematic design of the 
current ANL MCP photodetector assembly and its electrical circuit diagram.
 
The MCP photodetector is an all glass body assembly, consisting of a glass base 
window, a top window, two side walls, three grid spacers and two MCPs. The side 
walls are frit-bonded onto the base window, with silver stripline anodes 
printed to lead the signals and high voltage connections to the outside. Three 
grid spacer sets are used; one between the top window and the upper MCP, one 
between the two MCPs, and one between the lower MCP and the anode. These 
spacers function as insulators to separate the different components, and also 
to hold these internal components in place in the vacuum assembly. 

When light hits the photocathode, photoelectrons are emitted and are then 
accelerated towards the MCP hitting the inner wall of a channel and emitting 
secondary electrons. Then, the secondary electrons are further accelerated by 
the electric field generated across the ends of the MCP bombarding the channel 
wall again and producing additional secondary electrons.  This process is 
repeated as the electrons travel along the channels of the MCPs. Two MCPs are 
used in our design to ensure that the electrons are multiplied several times, 
and the signals deposited onto the stripline anode are strong enough for signal 
detection. 

\begin{figure}[tbp]
\centering \includegraphics[scale=0.23]{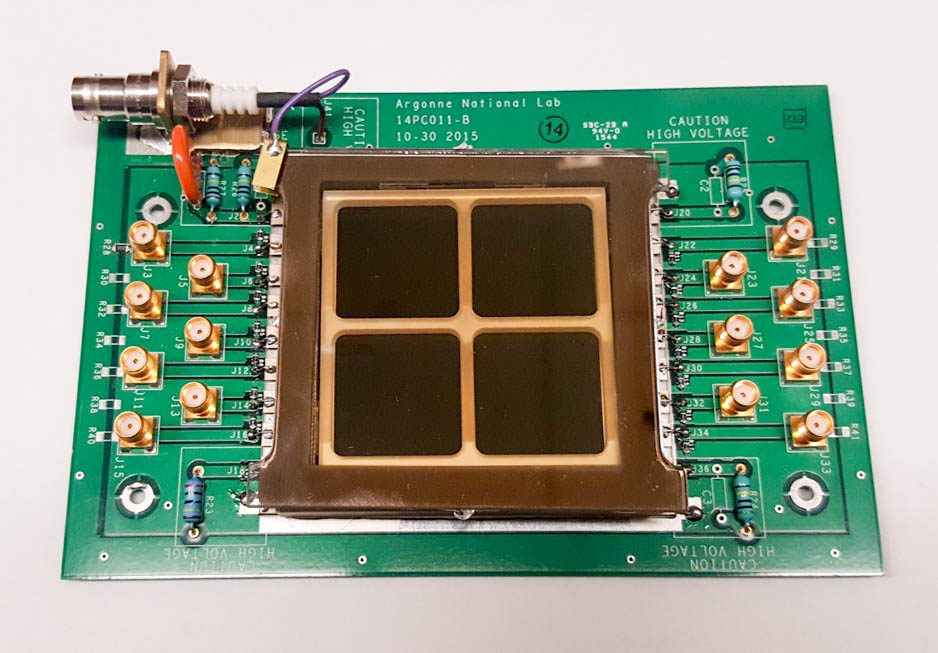}
\caption{Completed 6$\times$6~cm$^2$ MCP photodetector attached to a circuit 
board, providing firm electrical connections.} \label{fig:MCP_assm}
\end{figure}

Four ultra-thin metal shims are applied at the top and bottom surfaces of the 
two MCPs to lead the electrical connection to external connections. The 
detailed description of the circuit connection inside the vacuum package can be 
found in reference \cite{Xia-MCPs}. This independent bias-voltage design 
provides the advantage of individually controlling and fine tuning the bias 
voltage for each MCP. A Bialkali photocathode is deposited on the inner surface 
of the top window, and an indium seal is made between the top window and the 
side wall through a low temperature thermo-compression sealing process to form 
a hermetic vacuum detector package. The completed MCP photodetector is attached 
to a custom-made circuit board, providing a permanent mount and firm electrical 
connections, as shown in Figure \ref{fig:MCP_assm}. External electrical 
connections for both signal and high voltage are inserted into the external 
resistor connections to serve as high voltage dividers, ensuring both MCPs work 
at an independently optimized high voltage for the best performance.  
Additional capacitors may also be added across the resistor divider for better 
signal waveform.  

The microchannel plates used in the 6$\times$6~cm$^2$ MCP photodetector are 
diced from the next-generation large area (20$\times$20~cm$^2$) MCPs 
\cite{LAPPD,Craven-MCPs}, the world's largest commercially available MCPs.  
These next-generation MCPs are produced through a glass drawing process and 
functionalized through the ALD process. This is completely different from the 
production of traditional leaded glass MCPs. The glass drawing process uses 
borosilicate glass as the tube material, which is considerably less expensive 
than leaded glass and eliminates the chemical etching process required in the 
traditional method. This makes it much more cost-effective for MCP production.  
Here, we use standard borosilicate glass MCPs with 20~$\mu$m pore size, 60:1 
L/d (pore length to diameter) ratio and 8$^{\circ}$ bias angle relative to the 
MCP surface normal. The two MCPs are placed as the chevron configuration in the 
vacuum package, which reverses the bias angle to -8$^{\circ}$. 

\section{Rate capability measurement} \label{sec_proton_measurements}
The rate capability of MCP photodetectors is one of the most critical 
parameters for applications in high luminosity environments, such as an EIC.  
Due to the high resistive layer coating on ALD functionalized MCPs, the current 
in the MCP pores may not flow off fast enough when the MCP detector is exposed 
to high particle rates. This effect may cause severe charge saturation reducing 
the gain of the MCP photodetector, and limiting the detector performance. 
 
We investigated the rate capability of the 6$\times$6~cm$^2$ MCP photodetectors 
with the 120~GeV/c primary proton beam at Fermilab Test Beam Facility (FTBF).  
As the protons travel through the 3 mm thick glass top window of the MCP 
photodetector, \u Cerenkov light is produced and detected by the MCP 
photodetector as our signals. The beam is delivered as a slow spill with a 4~s 
duration once per minute with a maximum intensity of $~$10$^5$ particles per 
spill. The beam has an elliptical shape, a major-diameter of 4.6~mm and a 
minor-diameter of 1.8~mm, and a Gaussian density profile. The incident proton 
beam was monitored by an upstream multiwire proportional counter to see the 
beam profile. Three plastic scintillators were used in coincidence as a trigger 
and to count the number of incident protons. A light-tight dark box was 
designed to hold the MCP photodetector in the beam path with the detector 
surface facing the beam direction. High voltage was applied to the MCPs through 
an external resistor voltage divider, and signals from the striplines were read 
out through a DT5742 desktop digitizer \cite{Digitizer} produced by CAEN 
(Costruzioni Apparecchiature Elettroniche Nucleari S.p.A.) with a sampling rate 
of 5~GS/s. The digitizer is based on a switched capacitor array of DRS4 (Domino 
Ring Sampler) chips \cite{DRS}, 16 analog input channels, and one additional 
analog input for fast trigger.

\begin{figure}[tbp]
\centering \includegraphics[scale=0.25]{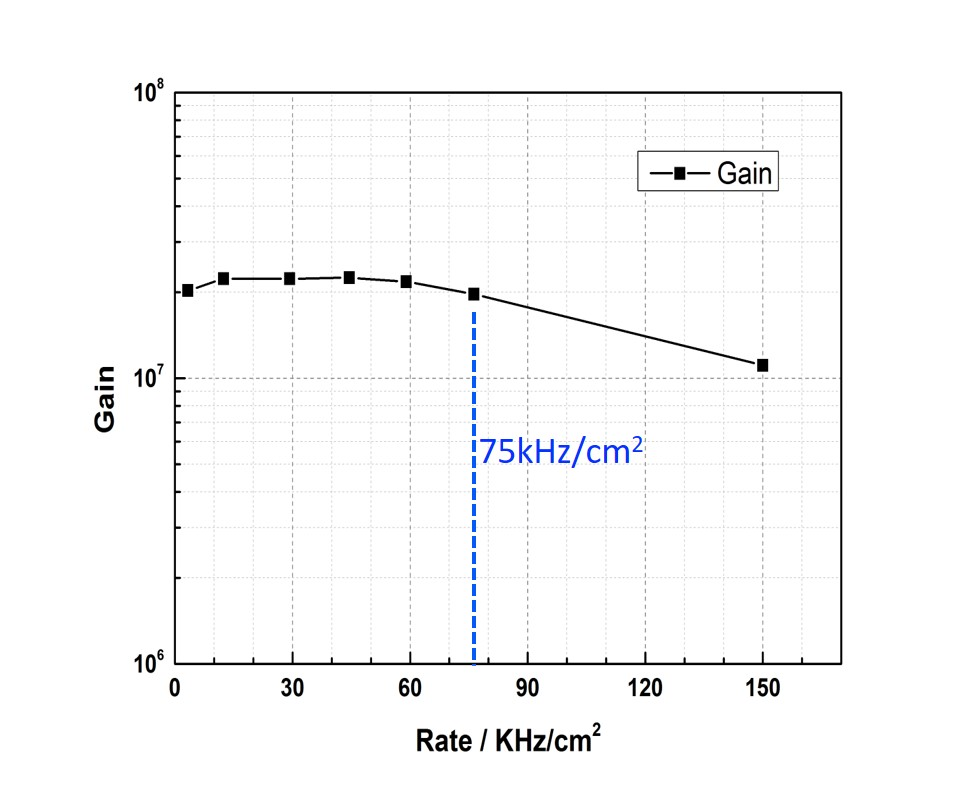}
\caption{Gain of the MCP photodetector as a function of the 120 GeV/c proton 
beam flux. The gain of the detector is stable up to beam flux of 75 kHz/cm$^2$, 
and the gain is still over 10$^7$ at 150 kHz/cm$^2$. } 
\label{fig:MCPs_gain_proton_beam}
\end{figure}

During our experiment, the 120~GeV/c proton beam intensity was tuned to vary 
from 500 to 40,000 particles per spill. The beam rate was calculated using the 
number of triggers per spill and was corrected for the size of the beam spot by 
reconstructing the beam profile. The calculated beam rate varied from 3 to 
150~kHz/cm$^2$ corresponding to the monitored beam particle intensity. Figure 
\ref{fig:MCPs_gain_proton_beam} shows the gain of the MCP photodetector 
measured as a function of the beam rate. The measured gain of the investigated 
detector is stable up to a beam flux of 75~kHz/cm$^2$, and is still over 10$^7$ 
when the beam flux reaches 150~kHz/cm$^2$. Such a high rate capability of the 
MCP photodetector will be sufficient for EIC PID detectors at the proposed beam 
energies.

\section{Magnetic field tolerance measurement}\label{sec_B_measurement}
In the EIC detector concepts, solenoid magnets with field strengths of 
1.5~Tesla are proposed. The imaging \u Cerenkov detectors (RICH, DIRC) and 
time-of-flight systems are designed to cover the area of the barrel and end 
caps for charged particle ($e$/$\pi$/$K$/$p$) separations. This compact design 
requires the photosensors to work properly in harsh environments with magnetic 
field strengths up to 1.5 Tesla. 

\begin{figure}[tbp]
\centering \includegraphics[scale=0.27]{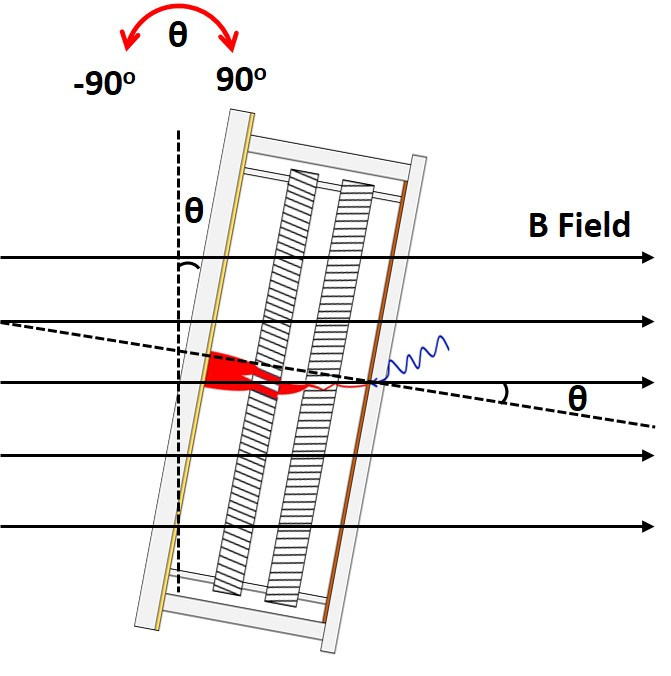}
\caption{Schematic of the rotation mechanism of the MCP photodetector with 
with angle $\theta$ relative to the magnetic field direction during the 
measurement.} \label{fig:MCPs_theta_rotation}
\end{figure}

At ANL, a decommissioned superconducting magnet from a magnetic resonance 
imaging (MRI) scanner was acquired to test instruments for the muon $g-2$ 
experiment \cite{Magnet}. The magnet provides a large bore with a diameter of 
68~cm and a very homogeneous field (7~ppb/cm), with a tunable strength of the 
magnetic field up to 4~Tesla. We have built a characterization system 
compatible with the solenoid magnet to test the performance of the 
6$\times$6~cm$^2$ MCP photodetector in a strong magnetic field environment.  
The MCP photodetector was fixed in a custom built non-magnetic, light-tight 
dark box. The dark box was held on a test platform with the detector surface 
normal to the direction of magnetic field. The position of the dark box was 
adjusted so that the center of the MCP photodetector was well-aligned with the 
center of the solenoid magnet. A rotation mechanism was also integrated with 
the system, allowing rotation of the MCP photodetector with an angle $\theta$, 
as shown in Figure \ref{fig:MCPs_theta_rotation}. A 405 nm light-emitting diode 
(LED) was used as the light source and was introduced to the surface of the MCP 
photodetector through an optical fiber. High voltage was applied to the MCPs 
from a supply with variable voltage control, and signals from the striplines 
were read out through the CAEN DT5742 desktop digitizer.

\subsection{Magnetic field strength dependence}\label{subsec_HV}
\begin{figure}[tbp]
   \hspace{-0.5cm} \includegraphics[scale=0.4]{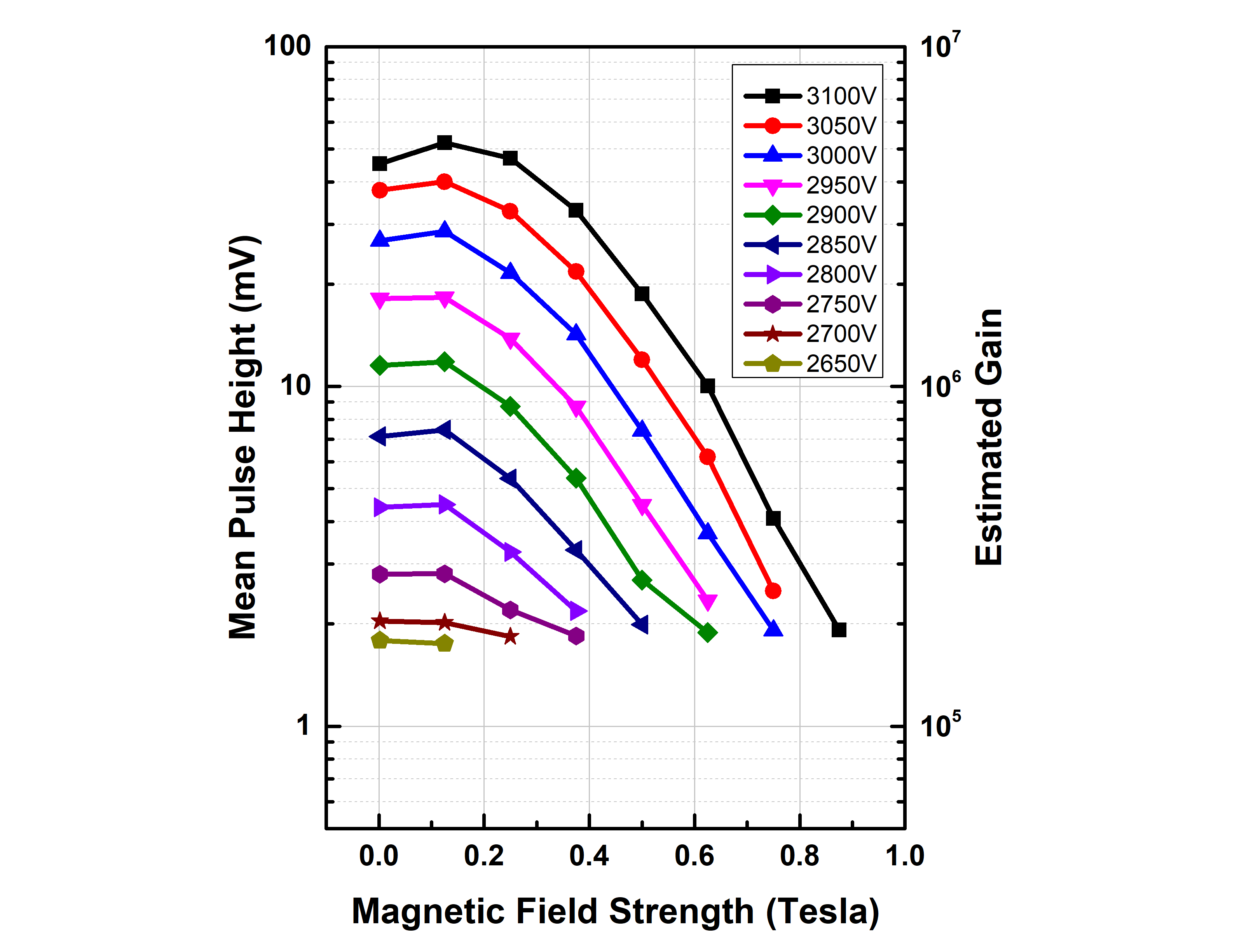}
\caption{Dependence of the MCP photodetector response, in terms of the mean 
pulse height, on the magnetic field strength at different high voltages.} 
\label{fig:MCPs_gain_B_HV}
\end{figure}

The dependence of the MCP photodetector performance on the magnetic field 
strength was done at a rotation angle $\theta$ = 0$^{\circ}$, i.e., where the 
direction of the magnetic field is normal to the surface of the MCP 
photodetector. We measured the performance of the investigated MCP 
photodetector in various magnetic field strengths at different bias high 
voltages. The results are presented in Figure \ref{fig:MCPs_gain_B_HV}. At a 
fixed bias high voltage, the mean pulse height of the MCP photodetector 
increases slightly as the magnetic field strength increases to 0.2~T, decreases 
as the magnetic field strength continues to increase, and eventually breaks 
down at a fixed magnetic field strength of $\sim$~0.8~T. In the same magnetic 
field environment, the mean pulse height of the MCP photodetector increases as 
the high voltage increases. This behavior is similar to our previous 
measurements of the MCP photodetectors without applying a magnetic field 
\cite{Wang-MCPs}.

\subsection{Magnetic field angle dependence}\label{subsec_theta}

\begin{figure}[tbp]
\centering \includegraphics[scale=0.35]{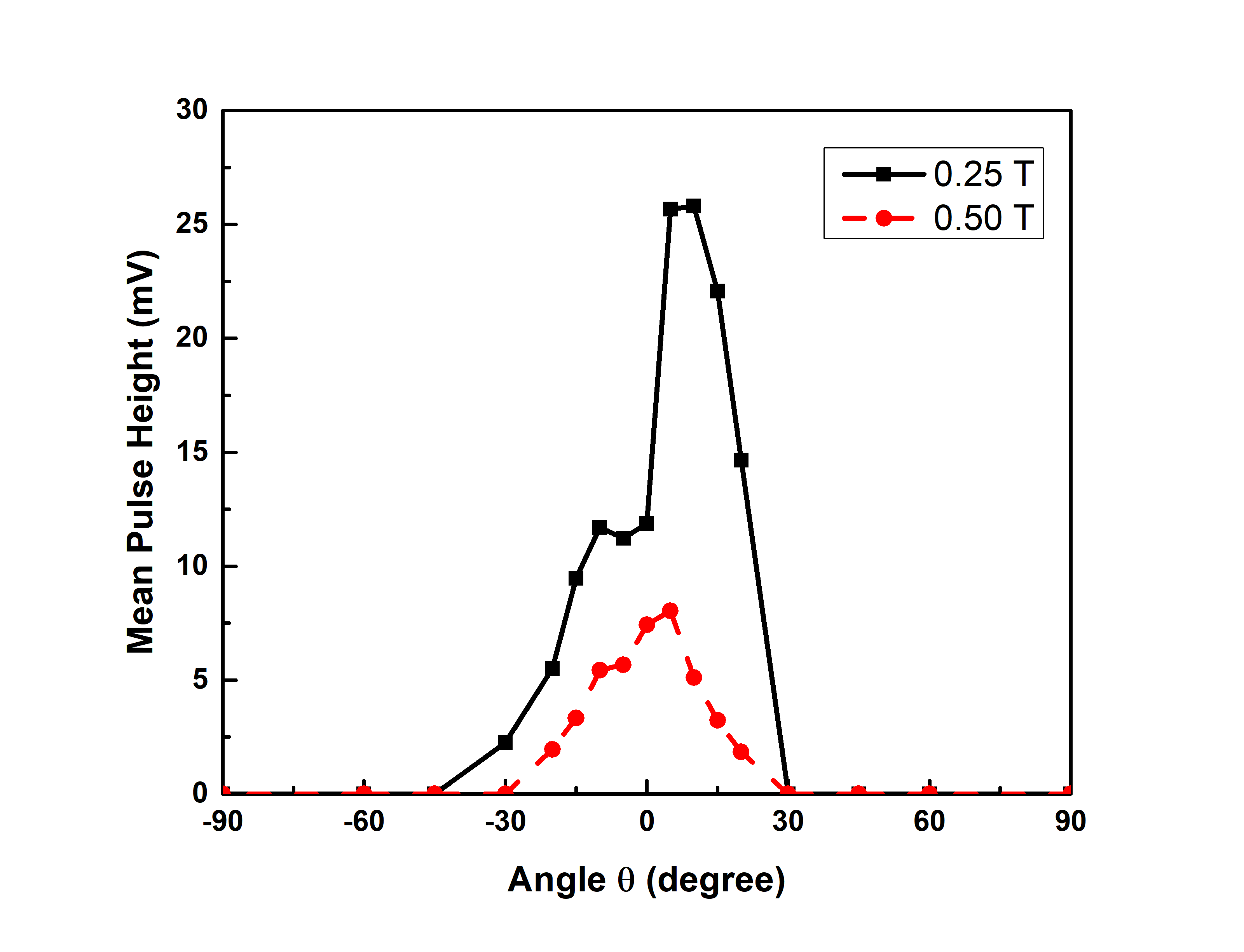}
\caption{Response of the MCP photodetector as a function of the rotation angle 
   $\theta$ relative to the direction of magnetic field. The two peaks around 
-8$^{\circ}$ and 8$^{\circ}$ indicates the effect due to the 8$^{\circ}$ bias 
angle of the MCPs. Note that the intensities of these two peaks are not the 
same due to the different effect from top and bottom MCPs.} 
\label{fig:MCPs_gain_theta_B}
\end{figure}
The angular dependence of the MCP photodetector was additionally studied by 
rotating the MCP photodetector along with angle $\theta$ relative to the 
magnetic field direction, as shown in Figure \ref{fig:MCPs_gain_B_HV}. We fixed 
the bias high voltage at 3000 V on the photodetector and rotated the 
photodetector from -90$^{\circ}$ to 90$^{\circ}$ for a full range of angle 
measurements. Figure \ref{fig:MCPs_gain_theta_B} presents the response of the 
MCP photodetector, in terms of the mean pulse height, measured as a function of 
the rotation angle $\theta$ at two magnetic field strengths of 0.25 and 
0.5~Tesla. For $\theta$~$\leq$~-30$^{\circ}$ or $\theta$~$\geq$~30$^{\circ}$, 
any signal is hardly seperated from the noise level. Within 
-30$^{\circ}$~$\leq$~$\theta$~$\leq$~30$^{\circ}$, there are two peaks at about 
$\theta$ = $\pm$8$^{\circ}$, which are due to the chevron configuration of two 
MCPs inside the photodetector. The detector's response reaches a maximum when 
the pore of either MCP is well-aligned with the magnetic field direction.  The 
intensities of the two peaks are different, which is due to the different 
effect from the top and bottom MCPs.

\subsection{Design optimization of MCP photodetector}\label{subsec_opt}
In the design of the EIC experiment, a 1.5~Tesla solenoid magnet will be used 
for tracking charged particles. The magnetic field tolerance requirement varies 
from detector to detector depending on their distances from the interaction 
point and the magnetic field direction. From our measurement, the 
6$\times$6~cm$^2$ MCP photodetector has shown a good magnetic field 
tolerance of up to 0.8 Tesla, comparable to that of current commercially 
available MCP-PMTs ($\sim$~1.0~T) with similar pore size \cite{MCPs-B}. Here, 
we must emphasize that the current LAPPD design is not yet optimized for 
magnetic field tolerant applications. The distances between the photocathode, 
MCPs, and the anode are relatively large in the LAPPD design. For instance, the 
spacing between the photocathode and the top MCP of 2~mm and spacing between 
the bottom MCP and the anode of 3.2~mm \cite{Wang-MCPs}. To optimize for a 
magnetic field environment, these distances should be reduced to minimize the 
electron transit distance. Meanwhile, MCP photodetectors with smaller pore 
sizes have shown better magnetic field tolerance than those with larger pore 
sizes \cite{MCPs-B, Lehmann, Ilieva}. A redesign of the current LAPPD 
configuration with smaller pore sizes (e.g.  10~$\mu$m or even 5~$\mu$m) and 
reduced distances between the PMT elements should improve its magnetic field 
tolerance. 

\section{Conclusions}
We have described the current design of 6$\times$6~cm$^2$ microchannel plate 
photodetectors with the next-generation MCPs functionalized through atomic 
layer deposition process. The rate capability and magnetic field tolerance of 
these photodetectors were tested using Fermilab's 120 GeV proton beam and 
Argonne's 4~Tesla magnetic field facility. The photodetectors exhibit stable 
performance for rates up to 75~kHz/cm$^2$ and magnetic field tolerance for 
fields up to 0.8~Tesla. The magnetic field angle dependence was also measured, 
showing enhanced performance at $\pm$8$^{\circ}$ tilt angle due to the original 
MCP 8$^{\circ}$ bias angle. The magnetic field tolerance of these detectors 
could be further improved by using smaller pore size MCPs and redesigning the 
package with reduced distances between the photocathode, the MCPs, and the 
anode.

\section{Acknowledgments}
The authors would like to thank Frank Skrzecz (Engineer at ANL) for his 
mechanical engineering support; Joe Gregar (Scientific Glass Blower at ANL) for 
his talented work on glass parts; Peter Winter (Physicist at ANL) for his 
arrangement of the Argonne magnetic facility usage; the staff at Fermilab Test 
Beam Facility for their beamline support; and many people from the LAPPD 
collaboration for their advices and assistants.  This material is based upon 
work supported by the U.S. Department of Energy, Office of Science, Office of 
High Energy Physics, Office of Nuclear Physics, under contract number 
DE-AC02-06CH11357. Work at Brookhaven National Laboratory was supported by the 
U. S. Department of Energy, Office of Science under contract No. 
DE-AC02-98-CH10886. This work was also partially supported by the EIC R{\&}D 
funding from the Office of Nuclear Physics and Office of Science of the U.S.  
Department of Energy.

\end{document}